\documentclass[onecolumn,preprintnumbers,11pt,amsmath,amssymb,nofootinbib]{revtex4}

\def	\be	{\begin{equation}}
\def	\ee	{\end{equation}}
\def	\bqt	{\begin{quote}}
\def	\eqt	{\end{quote}}
\def	\a	{\alpha}

\def	\mn 	{\mu \nu}
\def	\pl	{\partial}
\def	\del	{\nabla}

\begin{document}

\title{Derivation of the Null Energy Condition}

\author{Maulik Parikh$^{1}$ and Jan Pieter van der Schaar$^{2}$}

\affiliation{$^{1}$Department of Physics and Beyond: Center for Fundamental Concepts in Science \\
Arizona State University, Tempe, Arizona 85287, USA \\
$^{2}$Delta Institute for Theoretical Physics, Institute of Physics \\
University of Amsterdam, Science Park 904, 1098\, XH Amsterdam, The Netherlands} 

\begin{abstract}
\begin{center}
{\bf Abstract}
\end{center}
\noindent
We derive the null energy condition, understood as a constraint on the Einstein-frame Ricci tensor, from worldsheet string theory. For a closed bosonic string propagating in a curved geometry, the spacetime interpretation of the Virasoro constraint condition is precisely the null energy condition, to leading nontrivial order in the $\a'$ expansion. Thus the deepest origin of the null energy condition lies in worldsheet diffeomorphism invariance.
\end{abstract}

\maketitle

\section{Introduction}
\label{introduction}

\thispagestyle{empty}

\noindent
In the absence of physical constraints, any spacetime metric is formally an exact solution to Einstein's equations, as one can simply define the energy-momentum tensor to be equal to the Einstein tensor for the given metric. Many plausible physical constraints have been proposed, from cosmic censorship to global hyperbolicity, but perhaps the most important are the energy conditions. The energy conditions are local inequalities, traditionally expressed on the energy-momentum tensor, $T_{\mn}$, that, loosely speaking, capture the positivity of local energy density. The most fundamental of these, the null energy condition (NEC), requires that, at every point in spacetime,
\be
T_{\mn} v^\mu v^\nu \geq 0 \; ,	\label{TNEC}
\ee
for any light-like vector, $v^\mu$. In addition to its role in constraining the set of physical spacetimes, the NEC is invoked in several key gravitational theorems. For example, the NEC is a central assumption in the Hawking-Penrose singularity theorems \cite{Hawking:1969sw} that ensure the existence of the Big Bang singularity, as well as in the second law of thermodynamics for black holes \cite{Bardeen:1973gs}. 

Although the energy conditions may appear reasonable, they are nevertheless ad hoc physical constraints without a clear origin in fundamental theory \cite{Curiel}. Their validity has consequently been questioned \cite{Barcelo:2002,Urban:2010vr,Rubakov:2014}. Indeed, the strong energy condition has already been fruitfully discarded in inflationary cosmology. However, the null energy condition appears to be of a different character. In particular, its role in black hole thermodynamics should give us pause about dispensing with it; numerous thought experiments have shown that black hole thermodynamics is essential to the consistency of ordinary thermodynamics.

It is therefore important to prove the NEC from first principles. Now, expressed in the form of (\ref{TNEC}), the NEC appears as a property of the energy-momentum tensor for matter. One might think then, that some basic principle of quantum field theory (QFT), our extraordinarily powerful framework for describing matter, would ensure (\ref{TNEC}), perhaps in some appropriate classical limit. However, that does not appear to be the case. We now know of several effective field theories \cite{Rubakov:2014} --- such as theories of ghost condensates \cite{ArkaniHamed:2003} or conformal galileons \cite{Nicolis:2009} --- that violate (\ref{TNEC}) but seem perfectly consistent with the basic principles of QFT. Indeed, it is easy to come up with a semi-trivial counter-example. Consider the theory of a free massless ghost. The Lagrangian is merely that of a free massless scalar field but with the ``wrong" overall sign. An overall sign does not affect the classical equations of motion of course. In fact, the theory can be quantized as well. No instability of the vacuum arises since there is no coupling to other ``normal" particles with positive energy. And yet this theory violates the NEC; evidently, the NEC does not emerge readily from the principles of QFT.

To proceed, note that the NEC is typically invoked in the context of general relativity. Here the sign of the matter action {\em is} relevant because it is no longer an overall sign, but a sign relative to the Einstein-Hilbert action. Thus if we are to derive the NEC from first principles, we must do so in a theory that includes both matter and gravity. Moreover, the NEC is invariably used in conjunction with Einstein's equations. Then $T_{\mn} v^\mu v^\nu \geq 0 \Leftrightarrow R_{\mn} v^\mu v^\nu \geq 0$. Indeed, for the purpose of many theorems, the whole point of assuming (\ref{TNEC}) is to transmit the positivity condition to the Ricci tensor,
\be
R_{\mn} v^\mu v^\nu \geq 0 \; ,	 \label{Riccinec}
\ee
so that it can be put to use in Raychaudhuri's equation. Our first key insight, then, is to regard (\ref{Riccinec}) as the actual physical requirement; we will continue to refer to it as the null energy condition, even though, expressed in this form, it is now a statement about geometry rather than about energy densities. We will derive precisely this equivalent form of the null energy condition as a general consequence of worldsheet string theory. String theory of course is a theory of both matter and gravity and it is a promising place to look because it contains supergravity which, like most well-behaved theories, happens to satisfy the null energy condition as an accidental fact. The goals of this paper then are to derive the null energy condition from string theory, to thereby uncover the deepest origin of the NEC, and to identify the physical principle behind it.

\section{Spacetime Null Vectors from the Virasoro Constraints}
\label{NullVectors}

\noindent
Both versions of the null energy condition call for null vectors. Our first task, then, is to see how spacetime null vectors arise from worldsheet string theory. Consider a bosonic closed string propagating in flat space. We briefly recall some elementary facts about worldsheet string theory. The Polyakov action is
\be
S[X^\mu,h_{ab}] = - \frac{1}{4 \pi \a'} \int d^2 \sigma \sqrt{-h} h^{ab} \pl_a X^\mu \pl_b X^\nu \eta_{\mn} - \frac{c}{4 \pi} \int d^2 \sigma \sqrt{-h}  R \; .  \label{flat-action}
\ee
We will use Latin indices to denote the two worldsheet coordinates, which are taken to be dimensionless, and Greek indices for the $D$ spacetime coordinates. Here $h_{ab}$ is the dimensionless worldsheet metric and $R$ is the corresponding Ricci scalar. $c$ is an arbitrary dimensionless constant. The action is manifestly diffeomorphism- and Weyl-invariant. From the target-space perspective, $X^\mu(\tau, \sigma)$ are the spacetime coordinates of the worldsheet point labeled by $(\tau, \sigma)$. However, from the worldsheet perspective, the index $\mu$ is an internal index on a set of $D$ scalars. Thus, as the Lagrangian indicates, worldsheet bosonic string theory is simply a two-dimensional field theory of $D$ massless scalars, $X^\mu$, coupled to two-dimensional Einstein gravity.

The equation of motion for each $X^\mu$ is a wave equation:
\be
\left ( - \frac{\partial^2}{\partial \tau^2} +   \frac{\partial^2}{\partial \sigma^2} \right ) X^\mu (\tau, \sigma) = 0 \; .
\ee
This is subject to the periodic boundary condition $X^\mu(\tau, \sigma + 2\pi) = X^\mu (\tau, \sigma)$. The general solution is 
\be
X^\mu (\tau, \sigma) = x^\mu + \alpha^\prime \, p^\mu \tau + i \sqrt{\frac{\alpha^\prime}{2}} \,  \sum_{n \neq 0} \frac{\alpha^\mu_n}{n} e^{-i n (\tau - \sigma)} + i \sqrt{\frac{\alpha^\prime}{2}} \,\sum_{n \neq 0} \frac{\tilde \alpha^\mu_n}{n} e^{-i n (\tau + \sigma)} \; ,
\ee
where $x^\mu$, $p^\mu$, $\alpha^\mu_n$, and $\tilde \alpha^\mu_n$ are arbitrary constants, constrained only by a reality condition. Since the index $\mu$ just labels the worldsheet scalar, there is at this stage no particular reason for the spacetime vector $p^\mu$, say, formed by the collection of constants $p^0, ... p^{D-1}$ to be necessarily timelike, spacelike, or null.

The equations of motion for $h_{ab}$ are just Einstein's equations in two dimensions:
\be
-\frac{c}{2 \pi} \left (R_{ab} - \frac{1}{2} h_{ab} R \right ) = T_{ab} \; .
\ee
The left-hand side is zero because Einstein's tensor vanishes identically in two dimensions. The right-hand side is
\be
T_{ab} =  \frac{1}{2 \pi \a'} \left ( \pl_a X^\mu \pl_b X^\nu \eta_{\mn} - \frac{1}{2} h_{ab} (\pl X)^2 \right ) \; . \label{Tws}
\ee
Switching to light-cone coordinates on the worldsheet,
\be
\sigma^{\pm} \equiv \tau \pm \sigma \; ,
\ee
we can write the worldsheet metric locally as
\be
h_{ab} = \left ( \begin{array}{lc} \; \; 0 \; \;  \; -\frac{1}{2} \\ -\frac{1}{2} \; \; \;  \; \; \; 0 \end{array} \right ) \; ,
\ee
and we see that $T_{+-}$ is identically zero, even prior to using the equations of motion for $h_{ab}$. The remaining equations require
\be
T_{++} = 0 \quad , \quad T_{--} = 0 \; .
\ee
These are the Virasoro constraint conditions; as already mentioned, they are precisely Einstein's equations in two dimensions. Explicitly, from (\ref{Tws}), we have
\be
\pl_+ X^\mu \pl_+ X^\nu \eta_{\mn} = 0 \; ,
\ee
as well as a similar equation with $+$ replaced by $-$. Define a vector field $v_+^\mu = \pl_+ X^\mu(\sigma, \tau)$. Then
\be
\eta_{\mn} v_+^\mu v_+^\nu = 0 \; . \label{nullv}
\ee
That is, $v_+^\mu$ is null. We see that, through the Virasoro conditions, worldsheet string theory naturally singles out spacetime null vectors. 

\section{Null Energy Condition from the Virasoro Constraints}
\label{NEC-Virasoro}

\noindent
Now we make the obvious generalization to curved spacetime.  
Consider a closed bosonic string propagating on some background. We will not need to assume worldsheet conformal invariance at order $\a'$; we will not make use of any vanishing of the beta-functions for the background fields. 

The Polyakov action now reads
\be
S[X^\mu,h_{ab}] = - \frac{1}{4 \pi \a'} \int d^2 \sigma \sqrt{-h} h^{ab} \pl_a X^\mu \pl_b X^\nu g_{\mn}(X) - \frac{1}{4 \pi} \int d^2 \sigma \sqrt{-h} \Phi(X) R_h  \; . \label{curved-action}
\ee
We have replaced the Minkowski metric $\eta_{\mn}$ by the general spacetime metric $g_{\mn}(X)$. Consistent with worldsheet diffeomorphism-invariance, we have also allowed there to be a scalar field, $\Phi(X(\tau, \sigma))$, which is the dilaton field; we neglect the anti-symmetric Kalb-Ramond field, $B_{\mn}$, for simplicity. Also, we have added a subscript $h$ to the two-dimensional worldsheet Ricci scalar to distinguish it clearly from the curvature of $D$-dimensional spacetime.

We now perform a background field expansion $X^\mu(\tau, \sigma) = X^\mu_0 (\tau, \sigma) + Y^\mu (\tau, \sigma)$ where $X^\mu_0 (\tau, \sigma)$ is some solution of the classical equation of motion. Then, for every value of $(\tau, \sigma)$, we can use field redefinitions as usual \cite{CallanThorlacius,GSW1} to expand the metric in Riemann normal coordinates about the spacetime point $X^\mu_0 (\tau, \sigma)$:
\be
g_{\mn} (X) = \eta_{\mn} - \frac{1}{3} R_{\mu \a \nu \beta} (X_0) Y^\a Y^\beta - \frac{1}{6} \del_\rho R_{\mu \a \nu \beta} (X_0) Y^\rho Y^\a Y^\beta + ... \label{RNC}
\ee
Contracted with $\partial_a X^\mu \partial^a X^\nu$, the second and higher terms introduce quartic and higher terms in the Lagrangian; spacetime curvature turns (\ref{curved-action}) into an interacting theory:
\be
S_P[X,h] = S_P[X_0,h] - \frac{1}{4 \pi \a'} \int d^2 \sigma \sqrt{-h} h^{ab} \left (\eta_{\mu \nu} (\partial_a Y^\mu \partial_b Y^\nu) + R_{\mu \alpha \nu \beta}(X_0) \partial_a X_0^\mu \partial_b X_0^\nu Y^\alpha Y^\beta + ... \right ) \; .
\ee
The resultant divergences can be cancelled by adding suitable counter-terms to the original Lagrangian. Integrating out $Y$, the one-loop effective action is
\cite{CallanThorlacius,GSW1}
\be
S[X_0^\mu,h_{ab}] = - \frac{1}{4 \pi \a'} \int d^2 \sigma \sqrt{-h} h^{ab} \pl_a X_0^\mu \pl_b X_0^\nu (\eta_{\mn} + C_\epsilon \a' R_{\mn} (X_0)) -  \frac{1}{4 \pi} \int d^2 \sigma \sqrt{-h} C_\epsilon \Phi(X_0) R_h \; . \label{eff-action}
\ee
Here $C_\epsilon$ is the divergent coefficient of the counter-terms. 

In light-cone coordinates, the Virasoro constraints now read
\be
0 =  \pl_{\pm} X_0^\mu  \pl_{\pm} X_0^\nu \left ( \eta_{\mu \nu} + C_\epsilon \alpha' R_{\mu \nu} + 2 C_\epsilon\alpha' \nabla_\mu \nabla_\nu \Phi \right ) \; . \label{curvedVirasoro}
\ee
Defining $v_+^\mu = \pl_+ X^\mu$ as before, we recover (\ref{nullv}) at zeroeth order in $\a'$. Note that at every point $X^\mu_0 (\tau, \sigma)$ the spacetime metric $g_{\mn}(X)$ is just $\eta_{\mn}$; for a string passing through such a point, $v_+^\mu$ is therefore null with respect to both $\eta_{\mn}$ and $g_{\mn}(X)$. To derive the null energy condition, consider then an arbitrary null vector $v^\mu$ in the tangent plane of some arbitrary point in spacetime. Let there be a test string passing through the given point with either $\pl_+ X^\mu$ or $\pl_- X^\mu$ equal to $v^\mu$ at the point. By local Lorentz invariance, we can always find such a test string. Then, to first order in $\a'$, (\ref{curvedVirasoro}) says
\be
v^\mu v^\nu (R_{\mn} + 2 \nabla_\mu \nabla_\nu \Phi  ) = 0 \; .
\ee
This is tantalizingly close to our form of the null energy condition, (\ref{Riccinec}), but for two differences: it is an equality, rather than an inequality, and there is an additional, unwanted term involving the dilaton.

However, now we recall that the metric that appears in the worldsheet action is the string-frame metric. We can transform to Einstein frame by defining
\be
g_{\mn} = e^{\frac{4 \Phi}{D-2}} g_{\mn}^E \; .
\ee
Doing so, we find
\be
R^E_{\mn} v^\mu v^\nu  = + \frac{4}{D-2} (v^\mu \nabla^E_\mu \Phi )^2 \; .
\ee
The right-hand side is manifestly non-negative.   Hence we have
\be
R^E_{\mn} v^\mu v^\nu  \geq 0 \; .
\ee
This establishes the null energy condition. It is remarkable that the Virasoro constraints yield precisely the geometric form of the NEC, right down to the contractions with null vectors.

\section{Null Energy Condition in Lower Dimensions}
\label{NEC-lower dim}

\noindent
We have derived the null energy condition from the Virasoro constraint in bosonic string theory. 
The one-loop effective action we used was for a string moving in the critical number of spacetime dimensions i.e. $D = 26$ for a bosonic string. We could, though, just as well have considered a noncritical string moving in a lower-dimensional spacetime. In that case, the dilaton beta-function would include an additional term proportional to $D-26$ at the same order in $\a'$. However, the variation of this term with respect to $h^{ab}$ is proportional to $h_{ab}$ and so, since $h_{++} = h_{--} =0$, it does not contribute to the Virasoro constraints; our derivation of the NEC goes through as before. Hence we do not need to assume the critical number of dimensions, or indeed, more generally, that the background fields satisfy the vanishing beta-function constraints.

This already establishes the NEC in lower dimensions using worldsheet string theory. However, it is also interesting to derive the NEC for lower dimensions directly in field theory language \cite{Maldacena:2000}. An advantage of our geometric formulation of the NEC, (\ref{Riccinec}), is that it is easy to show that the NEC in lower dimensions follows from the NEC in higher dimensions. This is because, for a given compactification, it is straightforward to compute the lower-dimensional Ricci tensor in terms of the higher-dimensional Ricci tensor. Let us demonstrate this for compactification over a circle.

Consider first Kaluza-Klein reduction on a circle. Write the $D$-dimensional line element as
\be
ds^2 = \tilde{g}_{\ij}(x) dx^i dx^j + e^{2 \phi(x)} dy^2 \; .
\ee
Suppose there is a null vector $\tilde{v}^i$ in the $D-1$-dimensional space. Then there is an  associated null vector in the $D$-dimensional space, namely $v^\mu = (\tilde{v}^i, 0)$. Calculating the Ricci tensor of the $D$-dimensional metric and, transforming the lower-dimensional Ricci tensor, $\tilde{R}_{ij}$, to Einstein frame, yields
\be
\tilde{R}^E_{ij} \tilde v^i \tilde v^j = R_{\mn} v^\mu v^\nu + \left ( \frac{D-2}{D-3} \right ) (\tilde v^i \tilde \nabla^E_i \phi )^2 \; ,
\ee
where $\tilde{R}^E_{ij}$ and $\tilde \nabla^E_i$ refer to the lower-dimensional Einstein-frame metric. As the equation indicates, if $R_{\mn} v^\mu v^\nu \geq 0$ then $\tilde{R}^E_{ij} \tilde v^i \tilde v^j \geq 0$: the higher-dimensional NEC implies the lower-dimensional NEC.

Next, consider warped compactification. Write the line element as
\be
ds^2 = e^{2 A(y)} \tilde{g}_{\ij}(x) dx^i dx^j + dy^2 \; ,
\ee
where $A(y)$ is the warp factor. Calculating the Ricci tensors yields
\be
\tilde{R}_{ij} = R_{ij} + e^{2 A(y)} \tilde{g}_{ij} (A'' + (D-1) A'^2) \; ,
\ee
where $'$ denotes differentiation with respect to $y$. Again, consider a null vector $\tilde{v}^i$ in the $D-1$-dimensional space and its counterpart $v^\mu$ in the $D$-dimensional space.
Since $\tilde{v}$ is null, we have $\tilde{g}_{ij} \tilde{v}^i \tilde{v}^j = 0$. Hence
\be
\tilde{R}_{ij} \tilde{v}^i \tilde{v}^j = R_{\mn} v^\mu v^\nu \; ,
\ee
and, again, the validity of the $D$-dimensional NEC implies that of the $D-1$-dimensional NEC.

Thus, for both Kaluza-Klein and warped compactifications, if the NEC in $D$ dimensions holds, then the NEC in $D-1$ dimensions does. Since we have already established the null energy condition in the maximal (critical) number of dimensions, it follows by induction that the null energy condition applies in all lower dimensions, at least for toroidal compactifications. This, incidentally, is unlike the case for the weak energy condition, which is not inherited from higher dimensions.

\section{Discussion}
\label{Discussion}

\noindent
The null energy condition plays a vitally important role in gravity, in establishing the existence of the Big Bang singularity, in proving the second law of thermodynamics for black holes \cite{Bardeen:1973gs,Chatterjee:2012}, and in prohibiting the traversability of wormholes, the creation of laboratory universes \cite{Farhi:1987}, and the building of time machines \cite{Hawking:1992}. Moreover, it is the main local constraint that determines which solutions of Einstein's equations are physical. Here we have shown that the null energy condition, understood as a condition on the Ricci tensor, arises as the spacetime consequence of the Virasoro constraint in string theory.

Let us make a few observations. First, since we only used the Virasoro constraint and nowhere imposed the vanishing of the beta-functions, we never invoked the supergravity equations of motion explicitly. Second, in supergravity, as indeed in most well-behaved theories, the null energy condition happens to hold without any clear origin or principle associated with it. It is not evident from the supergravity Lagrangian alone whether the NEC had to hold because of some stability-related property of QFT, or because of supersymmetry (which is how the supergravity Lagrangian is constructed; note that the supersymmetry algebra implies that the Clifford vacuum has vanishing energy), or perhaps even because of black hole thermodynamics. By contrast, here we have identified the principle behind the NEC: it is reparameterization invariance --- worldsheet diffeomorphism invariance --- which gives rise to the Virasoro constraints whose spacetime interpretation is precisely the null energy condition, complete with contractions with null vectors. Moreover, we learn that the result has nothing to do with supersymmetry, having been obtained from the bosonic string. Third, simply because it is made up of the kinds of field content one usually encounters, the supergravity Lagrangian also obeys the weak and strong energy conditions. But these do not seem to emerge in any natural way from the worldsheet. And finally, let us just mention that, in holographic theories, the validity of the NEC in the bulk has been related to the existence of a c-theorem in the dual theory \cite{Freedman:1999gp}.

This work can be generalized in many directions. An obvious extension is to consider fermionic background fields in superstring theory. In string theory, the special role played by orientifolds, which do violate the null energy condition, is intriguing; it appears that they evade the implicit assumptions of our proof: orientifolds are end-of-the-world branes whereas the Riemann normal coordinate expansion used here must necessarily lie within an open set in spacetime. Also, having identified the origin of the NEC, we can see how it could in principle be extended by considering the Virasoro constraint in the presence of higher genus and $\a'$ corrections. The role of the worldsheet genus \cite{Fischler:1986} is especially interesting: in the field theory limit, higher genus corresponds to higher loops, where presumably the NEC in its usual form is violated, as we expect from considering quantum phenomena such as Casimir energy or Hawking radiation. It would also be very interesting to determine the Virasoro constraint at next order in $\a'$. This would indicate what the corresponding condition on the geometry is for higher-derivative gravity. Perhaps such an expression might help in obtaining the elusive proof of the second law of thermodynamics for black holes of higher-derivative gravity. 

In conclusion, we have found a very satisfying resolution of the origin of the null energy condition. As anticipated, the first-principles origin of the NEC lies not in the quantum field theory of matter, nor in general relativity, but in string theory, through an equation containing both matter and gravity. Moreover, the necessary condition can equally be regarded as being $R_{\mn} v^\mu v^\nu \geq 0$, a geometric equation. This constraint is the central premise in the proofs of a host of gravitational theorems. It is particularly pleasing that a constraint in spacetime is obtained on the worldsheet from the Virasoro constraint, itself a gravitational equation --- of two-dimensional gravity. Thus Einstein's equations in two dimensions restrict the physical solutions of Einstein's equations in spacetime. This then is another example of the beautiful interplay between equations on the worldsheet and in spacetime. On the worldsheet, a two-dimensional conformal field theory is coupled to two-dimensional Einstein gravity. In the presence of background fields, conformal invariance at order $\a'$ requires the vanishing of the beta-functions which, famously, yields Einstein's equations in spacetime, while, at the same order in $\a'$, the Virasoro constraint coming from worldsheet Einstein gravity is precisely the null energy condition in spacetime. 

\bigskip
\noindent
{\bf Acknowledgments}

\noindent
We would like to thank Jan de Boer, Joseph Polchinski, and Erik Verlinde for helpful discussions, and participants at the Amsterdam String Workshop for their feedback. M. P. is supported in part by DOE grant DE-FG02-09ER41624.


\bigskip

\begin{thebibliography}{99}


\bibitem{Hawking:1969sw} 
  S.~W.~Hawking and R.~Penrose,
  ``The Singularities of gravitational collapse and cosmology,''
  Proc. Roy. Soc. Lond. A {\bf 314}, 529 (1970).

\bibitem{Bardeen:1973gs} 
  J.~M.~Bardeen, B.~Carter, and S.~W.~Hawking,
  ``The Four laws of black hole mechanics,''
  Commun. Math. Phys.  {\bf 31}, 161 (1973).
  
\bibitem{Curiel}
E.~Curiel, ``A Primer on Energy Conditions," {\tt arXiv:1405.0403 [physics.hist-ph]}.
   
\bibitem{Barcelo:2002}
  C.~Barcelo and M.~Visser,
  ``Twilight for the Energy Conditions?,''
  Int. J. Mod. Phys. D {\bf 11}, 1553 (2002); {\tt gr-qc/0205066}.
  
\bibitem{Urban:2010vr} 
  D.~Urban and K.~D.~Olum,
  ``Spacetime Averaged Null Energy Condition,''
  Phys. Rev. D {\bf 81}, 124004 (2010); {\tt arXiv:1002.4689 [gr-qc]}.
  
\bibitem{Rubakov:2014} 
  V.~A.~Rubakov,
  ``The Null Energy Condition and its Violation,''
  Phys. Usp. {\bf 57}, 128 (2014); {\tt arXiv:1401.4024 [hep-th]}.
  
 \bibitem{ArkaniHamed:2003} 
  N.~Arkani-Hamed, H.~-C.~Cheng, M.~A.~Luty, and S.~Mukohyama,
  ``Ghost Condensation and a Consistent Infrared Modification of Gravity,''
  JHEP {\bf 0405}, 074 (2004); {\tt hep-th/0312099}.
  
\bibitem{Nicolis:2009}
A.~Nicolis, R.~Rattazzi, and E.~Trincherini, ``Energy's and Amplitude's Positivity," JHEP {\bf 1005}, 095 (2010);  {\tt arXiv:0912.4258 [hep-th]}.
  
\bibitem{CallanThorlacius}
C.~G.~Callan, Jr. and L.~Thorlacius, ``Sigma Models and String Theory," in {\em Particles, Strings and Supernovae: Proceedings of the Theoretical Advanced Study Institute in Elementary Particle Physics (TASI 88),} edited by A.~Jevicki and C.~I.~Tan (World Scientific, Teaneck, N. J., 1989), Vol. 2, 795.

  \bibitem{GSW1}
  M.~B.~Green, J.~H.~Schwarz, and E.~Witten,
 {\em Superstring Theory: Introduction} (Cambridge, London, 1987).
  
  \bibitem{Maldacena:2000} 
  J.~M.~Maldacena and C.~Nunez,
  ``Supergravity Description of Field Theories on Curved Manifolds and a No Go Theorem,''
  Int. J. Mod. Phys. A {\bf 16}, 822 (2001); {\tt hep-th/0007018}.
    
\bibitem{Chatterjee:2012} 
  S.~Chatterjee, D.~A.~Easson, and M.~Parikh,
  ``Energy Conditions in the Jordan Frame,''
  Class. Quant. Grav. {\bf 30}, 235031 (2013); {\tt arXiv:1212.6430 [gr-qc]}.
  
  \bibitem{Farhi:1987}
E.~Farhi and A.~H.~Guth, ``An Obstacle to Creating a Universe in the Laboratory," Phys. Lett. B {\bf 183}, 149 (1987).
  
\bibitem{Hawking:1992} 
  S.~W.~Hawking,
  ``The Chronology Protection Conjecture,''
  Phys. Rev. D {\bf 46}, 603 (1992).
  
\bibitem{Freedman:1999gp} 
  D.~Z.~Freedman, S.~S.~Gubser, K.~Pilch, and N.~P.~Warner,
  ``Renormalization group flows from holography supersymmetry and a c theorem,''
  Adv. Theor. Math. Phys. {\bf 3}, 363 (1999); {\tt hep-th/9904017}.
  
\bibitem{Fischler:1986}
 W.~Fischler and L.~Susskind,
``Dilaton Tadpoles, String Condensates and Scale Invariance II,''
Phys. Lett. B {\bf 173}, 262 (1986).

\end{thebibliography}
\end{document}